\documentclass[a4paper,11pt]{article}
\pdfoutput=1 % if your are submitting a pdflatex (i.e. if you have
\usepackage{jcappub} % for details on the use of the package, please
\usepackage[T1]{fontenc} % if needed
\usepackage{graphicx,amsmath,amsfonts,amssymb,dcolumn,epsfig,bm,float}
\usepackage{epsfig}%
\usepackage{widetable}
\usepackage{graphics}
\usepackage[margin=1.in]{geometry}
\usepackage{rotating}
\usepackage{silence}
\def\beq{\begin{equation}}
\def\eeq{\end{equation}}
\def\bea{\begin{eqnarray}}
\def\eea{\end{eqnarray}}

\title{\boldmath Constraints on fourth order gravity from binary pulsar and gravitational waves}

\author{Shreya Banerjee,}
\author{Sayantani Bera,}
\author{Srimanta Banerjee}
\author{and Tejinder P. Singh}

\affiliation{Tata Institute of Fundamental Research, Homi Bhabha Road, Mumbai 400005, India}

\emailAdd{shreya.banerjee@tifr.res.in}
\emailAdd{sayantani.bera@tifr.res.in}
\emailAdd{srimanta.banerjee@tifr.res.in}
\emailAdd{tpsingh@tifr.res.in}

\date{\today}

\abstract{We have earlier proposed a fourth order gravity model as a possible explanation for late time cosmic acceleration, and for flattened galaxy rotation curves. The model has a free length parameter whose value depends on the scale of the system under study (e.g. the whole Universe, a galaxy, or a compact binary pulsar). In the present work, we investigate the constraints imposed on the free model parameter by Hulse-Taylor binary pulsar data: periastron advance; and emission of gravitational waves and consequent period decay. It is shown that the model is consistent with these observations, provided the length parameter is bounded from above.\\

\vspace{.1in}
}

\begin{document}
\maketitle
\flushbottom

\section{Introduction}

It has been 100 years since Einstein's Theory of General Relativity (GR) predicted the existence of spacetime ripples i.e. gravitational waves \cite{einstein}. Their direct detection last year is a landmark in the history of GR \cite{ligo1,ligo2}. The detection of Hulse Taylor binary pulsar in the year 1975 served as the first indirect confirmation to this prediction \cite{hulse,weisberg}. Subsequently, several binary pulsars were detected and till date they remain one of the strongest sources of gravitational radiation. In spite of the tremendous success of GR in explaining several facets of the universe and astrophysical phenomena, several other observations like flattening of galaxy rotation curves and late time cosmic acceleration are yet to find a universally accepted explanation. Although the $\Lambda CDM$ model is the best fit to observations, the theoretical fine tuning issues with the cosmological constant, and the lack of direct detection of dark matter, leave room for improvement. Hence it is useful to investigate modified theories of gravity which can provide an alternative explanation to one or more such phenomena for which GR might not be necessarily adequate.

A large class of such models have been proposed in the literature; a few of them can adequately explain some of these phenomena. Models that are being extensively studied are MOND \cite{mond,mond1,mond2,mond3,mond4}, Scalar-Tensor-Vector theories \cite{stvg,stvg1,stvg2,nulambda,vlambda,vlambda1,teves}, f(R) and f(T) \cite{f(r),f(r)1,f(r)2,f(r)3,f(r)4,f(r)5,f(r)6,f(t),f(t)1,f(t)2,f(t)3} theories of gravity, Fourth Order Gravity (FOG) model \cite{priti}, Brane-World gravity models \cite{braneworld,braneworld1,braneworld2,braneworld3,braneworld4}, massive gravity models \cite{massive}, Nonsymmetric Gravity Theory (NGT) \cite{ngt} and the Metric-Skew-Tensor Gravity (MSTG) theory \cite{mstg}, amongst others. For a more detailed review on modified gravity models, one may refer to \cite{mogreview}.

Some of the above models have already been studied and constrained by strong-field tests of gravity such as gravitational waves emitted from binary pulsars \cite{bin1,capo,bin2,binary,binary1,binary2,binary3,binary4,binary5,binary6,will,will1}. Constraints come from the comparison of theoretical predictions and observational data for the orbital period decay of different binary pulsar systems. Further constraints are imposed through the study of periastron advance and Einstein time delay. Some of the models have also tried to put bounds on the speed of gravitational waves and the numerical value of the gravitational constant, through the above mentioned tests \cite{cG,cG1,cG2}.

In this paper, we attempt to test one such modified theory of gravity (FOG), proposed earlier by the authors in \cite{priti,priti1,shreya}, for gravitational waves emitted from binary pulsar systems. This model is a fourth order modification to the Einstein's field equations with a free length parameter $L$. Formulation of this fourth order gravity model is motivated by (but independent of) the problem of averaging of Einstein's equations \cite{aseem,aseem1}. The physics that might underlie such a model has been discussed in some detail in \cite{priti,shreya}. This model has already been studied for galactic and Hubble scales in \cite{priti,shreya}. In the galactic case, $L$ is of the order of the size of the galaxy and this modified gravity model implies Yukawa type corrections to the inverse square law, which can  explain  the non-Keplerian rotation curves as seen in observations, without invoking dark matter. In the cosmological case, choosing $L = c/H_0$ ($H_{0}$ is the present value of the Hubble constant),  allows the universe to enter into an accelerating phase in the present epoch, without invoking the presence of a cosmological constant. Hence, it provides a common explanation to galaxy rotation curves and cosmic acceleration. It turns out that while this model successfully explains the late time acceleration of the universe and galaxy rotation curves, it is also compatible with binary pulsar data. The analysis leading to this result is the subject of this paper. 

Compact binary systems are strong sources of gravitational radiation and provide an unprecedented opportunity to test the effects of gravity \cite{ref2}. The effect of gravitational waves on the pulsar timing and the orbital period can be used as indirect evidence of their existence \cite{1ref2}. In the present work, we study the fourth order field equations in the context of weak field gravity for Hulse-Taylor binary pulsar.

The main motivation of this paper is to pursue the Fourth Order Gravity model for a linearised weak field metric and match the results with observations. Sections \ref{vacuum} and \ref{source} are devoted to the derivation of the solution of modified Einstein's field equations in the weak-field limit for vacuum and in the presence of source respectively. The energy loss due to gravitational wave emission from the binary system also results in orbital period decay whose value can be obtained from observations. The relation between energy loss and period decay has been derived in Section \ref{energyloss}. In Section \ref{quadrupole}, we rewrite the expression for energy loss in terms of orbital and model parameters. The change in periastron advance has been discussed in Section \ref{periastron}. Lastly, in Section \ref{constraints}, we put bounds on $L$ using the above observations for Hulse-Taylor binary pulsar and show that this model is consistent with the observed data.

\section{Vacuum solution of linearized fourth order gravity}
\label{vacuum}
A fourth order modification to Einstein's field equations has been proposed earlier in \cite{priti,priti1,shreya} in order to explain galaxy rotation curves and cosmic acceleration. In this work, we investigate this model in the context of gravitational waves originating from binary pulsar systems.
Here, the Einstein field equations are modified by adding a term containing the fourth derivative of the metric tensor $g_{\mu\nu}$
\begin{equation}
R_{\mu\nu}-\frac12 g_{\mu\nu}R=\kappa T_{\mu\nu}+L^{2}R_{\mu\alpha\nu\beta}^{;\alpha\beta}
\end{equation}
where $\kappa =8\pi G$ (in units of c=1). $L$ is the length scale at which the modification to the field equations becomes important.

We were motivated to consider these field equations as they arise from averaging of quadrupolar inhomogeneities in general relativity \cite{Szekeres,Zalaletdinov}. However it turns out that it is interesting to consider these equations as a modified gravity in their own right, independent of any averaging process. The equations provide an interesting modification to the inverse square law in the Newtonian limit. It should be noted that these field equations do not arise from an action principle, and hence they should be treated as an effective theory in which the scale $L$ arises from an effective treatment of an underlying theory which does derive from an action.
In order to get the vacuum solutions, we set $T_{\mu\nu}=0$ and arrive at
\begin{equation}
R_{\mu\nu}-\frac12 g_{\mu\nu}R-L^{2}R_{\mu\alpha\nu\beta}^{;\alpha\beta}=0
\label{v1}
\end{equation}

The weak field static solutions for the modified field equations have already been studied extensively in \cite{priti}, which gives the biharmonic Poisson equation. In this work, we study the non-static weak field limit of the metric, which allows us to examine the properties of the gravitational waves.
 In order to find the wave equation satisfied by the gravitational wave, we perturb the metric upto $\mathcal{O}(1)$ \cite{hartle,Weinberg,Carroll} i.e. we set
\begin{equation}
g_{\mu\nu}=\eta_{\mu\nu}+h_{\mu\nu} \quad ({\rm where}\ h_{\mu\nu}\ll 1) 
\end{equation}
Here $\eta_{\mu\nu}$ is the Minkowski metric $(-1,+1,+1,+1)$ and $h_{\mu\nu}$ is the small perturbation to the flat Minkowski space-time.
Perturbing Eq. (\ref{v1}), we get
\begin{equation}
\delta R_{\mu\nu}+\frac{L^{2}}{2}\delta[g_{\mu\nu}g^{\rho\sigma}R_{\rho\gamma\sigma\delta}^{;\gamma\delta}]=L^{2}\delta[R_{\mu\alpha\nu\beta}^{;\alpha\beta}]
\end{equation}
The second term on the left hand side is the Ricci scalar term which is obtained by taking the trace of the field equation \eqref{v1}. In GR, the trace gives $R=0$, but in our case $R$ is non-zero.
To simplify the mathematical formulation, we choose the Lorentz gauge i.e. we set
\begin{equation}
\partial_{\nu}h_{\mu}^{\nu}-\frac12\partial_{\mu}h=0
\end{equation}
In this gauge, the perturbed field equations become
\begin{equation}
\Box h_{\mu\nu}-\eta_{\mu\nu}\partial_{\mu}\partial_{\nu}h^{\mu\nu}+\eta_{\mu\nu}\Box h+\frac{L^{2}}{2}\eta_{\mu\nu}\partial^{\alpha}\partial^{\beta}\Box h_{\alpha\beta}=L^{2}\left(\Box^{2} h_{\mu\nu}-\partial^{\beta}\partial_{\nu}\Box h_{\mu\beta}-\partial^{\alpha}\partial_{\mu}\Box h_{\alpha\nu}+\partial^{\alpha}\partial^{\beta}\partial_{\mu}\partial_{\nu}h_{\alpha\beta}\right)
\label{perturb}
\end{equation}
These equations become simpler if we apply the TT (traceless and transverse) gauge which is the general convention used to study gravitational waves. The TT gauge is given by
\begin{equation}
h_{\mu}^{\mu}=0 \quad ; \ \partial^{\nu}h_{\mu\nu}=0
\end{equation}
This gauge reflects the transverse nature of the gravitational waves.
Applying TT gauge, we get the modified wave equations (free space) as
\begin{equation}
(L^{2}\Box^{2}-\Box)h_{\mu\nu}=0
\label{v2.2}
\end{equation}
where $\Box=\eta_{\mu\nu}\partial^{\mu}\partial^{\nu}$ is the D'Alembertian operator. The first term inside the bracket gives the fourth order modification to the free space wave equation. 

In the following sections, we will study the physical significance of the solution of the modified wave equations.

\section{Modified wave equation in the presence of a source}
\label{source}

For analysing binary systems, we need to study gravitational radiation in the presence of sources.
Following the same procedure as for the vacuum case, we get the modified linearized Einstein's equation in TT gauge as
 \begin{equation}
 \Box\left(\Box-\frac{1}{L^2}\right)h_{\mu\nu}=\frac{16\pi G}{L^2} T_{\mu\nu}
 \label{MEE}
 \end{equation}
 The solution of the above equation can be obtained using a Green's function. Here we will consider that the source is compact, and  located in a region ${\bf x'}$ which includes the origin ${\bf x'}=0$,  and the observer is far away, at ${\bf r} = {\bf x}$. Throughout our calculations, we will restrict ourselves to slowly moving, compact sources \cite{hartle,Weinberg,Carroll} and because the system is compact it follows that  $|{\bf x'}| \ll |{\bf x}|$, which implies that most radiation is emitted at frequencies such that $r \gg 1/\omega$. Hence, we can take ${\bf r} \approx {\bf x}-{\bf x}'$.
 
The equation satisfied by the Green's function $G(r,t-t')$ is,
\begin{equation}
\Box\left(\Box-\frac{1}{L^2}\right)G(r,t-t')=4\pi\delta(t-t')\delta^{3}(\bf{r})
\label{s1}
\end{equation}
where $r=|\bf{x}-\bf{x'}|$ and $\delta(\bf{r})$ denotes the Dirac delta function. The Green's function for a similar fourth order equation has been calculated in \cite{Greens} (see Eq. 49 in Sec III D of \cite{Greens} and Appendix of their paper for details of the calculation). Following their approach, the corresponding retarded Green's function in our case can be shown to be
\begin{equation}
G(r,t-t')=\frac{L\mathcal{J}_1(\sqrt{(t-t')^2-r^2}/L)}{\sqrt{(t-t')^2-r^2}}\Theta(t-t'-r)
\label{green}
\end{equation}
where $\mathcal{J}_n$ is the Bessel function of the first kind and $\Theta$ is the Heaviside step function.
%We get back the Green's function for the unmodified wave equation when $L\to 0$ (GR limit) \cite{Greens}.
The general solution for the gravitational wave can be obtained by integrating the Green's function over all the sources
\begin{equation}
h_{\mu\nu}({\bf{x}},t)=\frac{4G}{L^2}\int d^{3}x'dt'G(r,t-t')T_{\mu\nu}({\bf{x}}',t')
\label{s6}
\end{equation}
Using Eq. (\ref{green}), we integrate Eq. (\ref{s6}) over $t'$ and ${\bf x'}$ to get
\begin{equation}
h_{\mu\nu}({\bf x},t)=\frac{4G}{L}\int\int d{\bf x'}\ dt'\ \frac{\mathcal{J}_1(\sqrt{(t-t')^2-r^2}/L)}{\sqrt{(t-t')^2-r^2}}\Theta(t-t'-r)T_{\mu\nu}(t',x')
\end{equation}
Now, as a consequence of the Bianchi identities,  the energy conservation equations for the modified field equations  are given by
\begin{equation}
(\kappa T_{\mu\nu}+L^{2}R_{\mu\alpha\nu\beta}^{;\alpha\beta})^{;\nu}=0
\end{equation}
We now perturb the above equations in the weak field limit (upto first order) and use the form of $\delta R_{\mu\alpha\nu\beta}^{;\alpha\beta}$ from  the right hand side of \eqref{perturb}. Writing the covariant derivative in terms of the partial derivative and Christoffel symbols, we neglect the terms involving Christoffel symbols as they produce higher order corrections. Thus retaining only the partial derivative, we get the energy conservation equations as
\begin{equation}
(\kappa \delta T_{\mu\nu}+L^{2}\delta R_{\mu\alpha\nu\beta}^{;\alpha\beta})^{,\nu}=0
\label{s9}
\end{equation}
Since we are considering weak field limit, the magnitude of energy momentum tensor must be small. Hence we are ignoring the higher order corrections to $T_{\mu\nu}$ such that the lowest non-vanishing term is same as the order of magnitude as the perturbation. Thus, henceforth, we will denote $\delta T_{\mu\nu}$ as $T_{\mu\nu}$ \cite{Carroll}. 

Taking the partial derivative of the right hand side of \eqref{perturb} (which denotes $\delta R_{\mu\alpha\nu\beta}^{;\alpha\beta}$) and applying Lorentz and TT gauge, it can be easily seen that $(\delta R_{\mu\alpha\nu\beta}^{;\alpha\beta})^{,\nu}=0$. Using this result and differentiating Eq. (\ref{s9}) with respect to time, we obtain \cite{hartle}
\begin{equation}
\frac{\partial^{2}}{\partial t_{r}^{2}}T_{00}=-\frac{\partial^{2}}{\partial x'_{i}\partial t_{r}}T_{0i}=\frac{\partial^{2}}{\partial x'_{i}\partial x'_{j}}T_{ij}
\end{equation}
Multiplying the above equation with $x'_{i}x'_{j}$ and integrating gives,
\begin{equation}
\frac{d^{2}}{dt_{r}^{2}}\int d^{3}x' x'_{i}x'_{j}T_{00}=\int d^{3}x'  x'_{i}x'_{j}\frac{\partial^{2}}{\partial x'_{k}x'_{l}}T_{kl}=2\int d^{3}x' T_{ij}
\label{s10}
\end{equation}
The quadrupole moment of the source is defined in terms of the energy momentum tensor as
\begin{equation}
I_{ij}(t')=\int d^{3}x' x'_{i}x'_{j}T_{00}(t',{\bf x}')
\end{equation}
Since in the chosen gauge, $h_{ij}$ is traceless and transverse, it is more convenient to replace $I_{ij}$ by its traceless (reduced) quadrupole moment defined as
\begin{equation}
Q_{ij}(t')=\int d^{3}x'(x'_{i}x'_{j}-\frac13\delta_{ij}r'^{2})T_{00}(t',{\bf x}')
\label{s11}
\end{equation}
In order to make $Q_{ij}$ transverse, we project its components on a transversal plane using the projection operator $P_{b}^{a}(x')=\delta_{b}^{a}-\frac{x'^{a}x'_{b}}{r'^{2}}$, where ${\bf x}'=(x',y',z')$ and $r'=|{\bf x}'|$. Using the basic properties of the projection operator i.e. $P^{2}=P$ and $P_{a}^{b}P_{c}^{a}=P_{c}^{b}$, we get the traceless-transverse quadrupole moment as
\begin{equation}
\bar{Q}_{ij}=P_{i}^{a}Q^{ab}P_{j}^{b}-\frac12 P^{ab}Q_{ab}P_{ij}
\label{s11.1}
\end{equation}
Using the above relations, we get the quadrupole formula for the emission of gravitational waves as 
\begin{equation}
\begin{split}
h_{ij}&=-\frac{2G}{L}\int_{-\infty}^{t-r/c}\frac{\mathcal{J}_{1}\left(\frac{\sqrt{\tau^2-r^2}}{L}\right)}{\sqrt{\tau^2-r^2}}\ddot{\bar{Q}}_{ij}(t')dt' \\
&=\frac{2G}{L}\int_{0}^{\infty}\frac{\mathcal{J}_{1}(s)}{\sqrt{s^2+\chi^2}}\ddot{\bar{Q}}_{ij}(t'_r)ds
\end{split}
\label{s12}
\end{equation}
Here $\tau=t-t'$, $\chi=r/L$, $s=c\sqrt{\tau^2-r^2/c^2}/L$ and $t'_r=\left(t-\frac{L\sqrt{s^2+\chi^2}}{c}\right)$.
 
 In the following section, we will use this expression to find the energy carried away by the gravitational waves emitted from a binary pulsar system.

\section{Energy Loss due to Gravitational Wave Emission}
\label{energyloss}

In this section, we shall investigate the energy carried by the gravitational wave. We consider the gravity wave solution in the far-field limit (i.e. ${\bf x} \gg {\bf x}' $). 
We have already seen that the weak field limit of the Einstein's equations (linear in the metric perturbation $h$) generates the gravitational wave. In order to find the energy carried by the gravitational wave, we have to go to the next higher order perturbation in $g_{\mu\nu}$ \cite{Carroll, mtw} i.e.
\begin{equation}
g_{\mu\nu}= \eta_{\mu\nu}+h_{\mu\nu} + h^{(2)}_{\mu\nu} 
\end{equation}
where $h^{(2)}_{\mu\nu} $ is the second order perturbation in $h_{\mu\nu}$.
The Einstein's field equations are satisfied at each order of perturbation \cite{Carroll}. Eq. $\eqref{MEE}$ represents the first order modified Einstein's equations. The second order modified Einstein's equations are given by 
\begin{equation}
G^{(1)}_{\mu\nu}[\eta+h^{(2)}] - L^2 (\nabla^{\alpha}\nabla^{\beta}R_{\mu\alpha\nu\beta})^{(1)}[\eta + h^{(2)}]  + G^{(2)}_{\mu\nu}[\eta+h] - L^2 (\nabla^{\alpha}\nabla^{\beta}R_{\mu\alpha\nu\beta})^{(2)}[\eta + h] = 0
\label{e01}
\end{equation}
where $(..)^{(1)}$ are the terms that are first order in $h^{(2)}$ and $(..)^{(2)}$ are the terms that are quadratic in $h$.

We can rewrite Eq. $\eqref{e01}$ in another form as
\begin{equation}
G^{(1)}_{\mu\nu}[\eta+h^{(2)}] - L^2 (\nabla^{\alpha}\nabla^{\beta}R_{\mu\alpha\nu\beta})^{(1)}[\eta + h^{(2)}] = \kappa t_{\mu\nu}
\end{equation}
where we have defined
\begin{equation}
\kappa t_{\mu\nu}= -\left(G^{(2)}_{\mu\nu}[\eta+h] - L^2 (\nabla^{\alpha}\nabla^{\beta}R_{\mu\alpha\nu\beta})^{(2)}[\eta + h] \right)
\label{e02}
\end{equation}

 %From a mathematical point of view, we again start by writing the full Einstein's equation in the context of our modified gravity theory 

%\begin{equation}
%R_{\mu\nu}-\frac{1}{2}g_{\mu\nu} R - L^2 \nabla^{\alpha}\nabla^{\beta}R_{\mu\alpha\nu\beta} = \kappa T_{\mu\nu}
%\label{e1}
%\end{equation}
%Here $T_{\mu\nu}$ is the matter stress-energy tensor.

%Adding the first order perturbation terms to both sides and taking all the exact terms to the right hand side, we can write,
%\begin{eqnarray}
%R^{(1)}_{\mu\nu}-\frac{1}{2}\eta_{\mu\nu} R^{(1)} - L^2 (\nabla^{\alpha}\nabla^{\beta}R_{\mu\alpha\nu\beta})^{(1)} && = \kappa T_{\mu\nu} +[-R_{\mu\nu}+\frac{1}{2}g_{\mu\nu} R + L^2 \nabla^{\alpha}\nabla^{\beta}R_{\mu\alpha\nu\beta} \nonumber \\  
%&& + R^{(1)}_{\mu\nu}-\frac{1}{2}\eta_{\mu\nu} R^{(1)}-  L^2 (\nabla^{\alpha}\nabla^{\beta}R_{\mu\alpha\nu\beta})^{(1)}]
%\end{eqnarray}
%By defining the term in the square brackets in the right hand side of the above equation as $t_{\mu\nu}$, we can rewrite the equation in the following form :
%\begin{equation}
%R^{(1)}_{\mu\nu}-\frac{1}{2}\eta_{\mu\nu} R^{(1)}= \kappa T_{\mu\nu} + \kappa t_{\mu\nu}
%\label{e3}
%\end{equation}
%where we have used,
%\begin{equation}
%\kappa t_{\mu\nu}= -R_{\mu\nu}+\frac{1}{2}g_{\mu\nu} R + L^2 \nabla^{\alpha}\nabla^{\beta}R_{\mu\alpha\nu\beta}+ R^{(1)}_{\mu\nu}-\frac{1}{2}\eta_{\mu\nu} R^{(1)}- L^2 (\nabla^{\alpha}\nabla^{\beta}R_{\mu\alpha\nu\beta})^{(1)}
%\label{e4}
%\end{equation}
%Equation $\eqref{e3}$ shows that the total contribution to the source term comes from the matter as well as the gravitational wave stress-energy.

One can, in principle, expand the exact Ricci tensor $R_{\mu\nu}$ in the above equation upto any desired order in $h$. Expansion upto the second order gives the energy and momentum carried by the gravitational wave and the successive higher orders represent the self-interaction of the gravitational waves and its contribution to the total energy-momentum. Here we are interested in only the next higher order i.e. the second order terms in the Einstein's equation. 

Since the gravitational wave is not localizable, we need to do an averaging over a macroscopic region (i.e. over a region large compared to the wavelength concerned) to get the effective energy-momentum as, 
\begin{equation}
\begin{split}
\langle t_{\mu\nu}\rangle= -\frac{1}{\kappa}\Big\langle\Big(-\frac14(\partial_{\mu}h_{\rho\sigma})(\partial_{\nu}h^{\rho\sigma})-\frac{1}{2}\eta^{\rho\lambda}(\Box h_{\rho\nu})h_{\lambda\mu}&+\frac{L^2}{2}\eta_{\mu\nu}\eta^{\delta\omega}\left(-\frac14(\Box^2h_{\delta\sigma})h^{\sigma}_{\omega}-\frac14 h_{\sigma\beta,\alpha\delta}h^{\beta\sigma,\alpha}_{,\omega}\right)\\
&+\frac{L^2}{4}\left((\Box^2 h_{\nu\sigma})h^\sigma_\mu+h_{\sigma\beta,\nu\alpha}h^{\beta\sigma,\alpha}_{,\mu}\right)\Big)\Big\rangle
\end{split}
\label{e13}
\end{equation}
The detailed steps to obtain Eq. $\eqref{e13}$ from Eq. $\eqref{e02}$ are given in Appendix \ref{A}.
 
We can use Eq. $\eqref{e13}$ to calculate the total energy radiated from a sphere to infinity. The energy flux in a particular direction $n^i$ is given by $\langle t_{0i} \rangle$, where $n^i$ is a unit spacelike vector normal to the surface of the sphere. The rate at which energy is radiated out from the sphere will be \cite{Carroll,mtw}
\begin{equation}
\begin{split}
\dot{E} &=  \int \langle t_{0i} \rangle n^i d^3 x \\
& =\int \langle t_{0r} \rangle n^r r^2 d\Omega \\
\end{split}
\label{el1}
\end{equation}
Substituting for $\langle t_{0r} \rangle$ from Eq. $\eqref{e13}$ we get,
\begin{equation}
\dot{E} = \int \frac{1}{4\kappa}\langle \langle \partial_0 h_{rr}\partial_r h^{rr}-L^2(h_{rr,rr}h^{rr,r}_{,0}+h_{rr,0r}h^{rr,0}_{,0})\rangle r^2 d\Omega 
\label{el2}
\end{equation}
Thus, the energy radiated by the source can be obtained by inserting Eq. $\eqref{s12}$ into Eq. $\eqref{el2}$. 
Substituting for the corresponding derivatives in the above equation and solving the integrals in the far-field limit i.e. $\chi \to \infty$, we get,
\begin{eqnarray}
\dot{E} &&= \int d\Omega \frac{1}{4\kappa}\Big\langle -4G^2\int_0^\infty \frac{\chi \mathcal{J}_1(s)}{\sqrt{s^2+\chi^2}}ds \int_0^\infty \frac{\chi^2\mathcal{J}_1(s)}{s^2+\chi^2}ds \dddot{\bar{Q}}_{rr}(t'_{r})\dddot{\bar{Q}}^{rr}(t'_{r}) \nonumber \\ 
&&~~~~~~~~~~~~~~~~~ - L^2\Big(-4G^2\int_0^\infty \frac{\chi\mathcal{J}_1(s)}{s^2+\chi^2}ds \int_0^\infty \frac{\chi^4\mathcal{J}_1(s)}{(s^2+\chi^2)^{3/2}}ds \ddddot{\bar{Q}}_{rr}(t'_{r})\ddddot{\bar{Q}}^{rr}(t'_{r}) \nonumber \\&&~~~~~~~~~~~~~~~~~~~~~~~~~ +4G^2\int_0^\infty \frac{\chi^3\mathcal{J}_1(s)}{s^2+\chi^2}ds \int_0^\infty \frac{\mathcal{J}_1(s)}{\sqrt{s^2+\chi^2}}ds \ddddot{\bar{Q}}_{rr}(t'_{r})\ddddot{\bar{Q}}^{rr}(t'_{r})\Big)\Big\rangle \\
&& = -4G^2\int d\Omega \frac{1}{4\kappa}\Big\langle (\dddot{\bar{Q}}_{rr}\dddot{\bar{Q}}^{rr}+2L^2\ddddot{\bar{Q}}_{rr}\ddddot{\bar{Q}}^{rr})\Big\rangle \nonumber \\
&& = -\frac{G}{5}\Big\langle (\dddot{Q}_{rr}\dddot{Q}^{rr}+2L^2\ddddot{Q}_{rr}\ddddot{Q}^{rr})\Big\rangle
\label{e10}
\end{eqnarray}
(The detailed derivation of the above form is given in Appendix \ref{B}.)
Inserting $c$ back  in the above equation we obtain,
\begin{equation}
\dot{E}=-\frac{G}{5c^5}\Big\langle (\dddot{Q}_{rr}\dddot{Q}^{rr}+\frac{2L^2}{c^2}\ddddot{Q}_{rr}\ddddot{Q}^{rr})\Big\rangle
\label{el11}
\end{equation}
We  observe that the energy loss now depends on both third and fourth time derivatives of the quadrupole moment, unlike in GR, where the fourth derivative in time is absent ($L=0$ in this case). Comparison of this loss rate against observations serves to constrain the model parameter $L$. 

\section{Quadrupole radiation from binary pulsar}
\label{quadrupole}
In this section we will consider a binary pulsar system as the source of gravitational radiation. We will express Eq. $\eqref{el11}$ in terms of the parameters of the  system. 

As we know, binary pulsars are strong sources of gravitational radiation. By measuring changes in the pulse repeating frequency and/or orbital period, one can determine the properties of the system needed to test the emission of gravitational radiation. In the present work, we shall consider the change in the orbital period only.
\begin{figure}[H]
\centering
  \includegraphics[width=10cm,height=6cm] {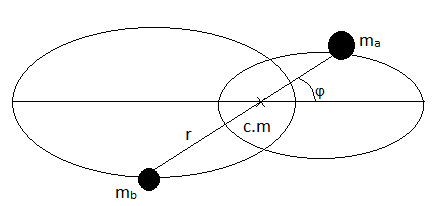}
\caption[Optional caption for list of figures]{ Schematic representation of a binary pulsar system with masses $m_a$ and $m_b$. Here $\tilde{r}$ is the distance between the two objects and $\phi$ is the angle subtended by the object at the centre of mass (c. m.) with respect to the semi-major axis.} 
\label{orbit}
\end{figure}
Let us consider a binary pulsar system in the centre of mass frame shown in Fig. \ref{orbit}. The system consists of two compact objects (may be white dwarf or neutron star), a and b, moving in an elliptic orbit with eccentricity $e$, orbital period $P$ and semi-major axis $a$. One or both of them may be pulsating. Some of the important parameters of the system have been described in Fig. \ref{orbit}. Below we have stated some other parameters which will be helpful in simplifying our calculations.
\begin{eqnarray}
{\rm Quadrupole\ moment}&=&I_{ij}(t')=x'_{i}x'_{j}\rho dV \nonumber \\
{\rm Position\ of\ first\ object\ with\ respect\ to\ c. m.}&=& x'_{i(a)}=\tilde{r}(\phi)\frac{m}{m_{a}}(\cos(\phi),\sin(\phi),0) \nonumber \\
{\rm Position\ of\ second\ object\ with\ respect\ to\ c. m.} &=& x'_{i(b)}=\tilde{r}(\phi)\frac{m}{m_{b}}(-\cos(\phi),-\sin(\phi),0) \nonumber \\
{\rm Total\ density}&=& \rho =\delta(x'^{3})[m_{a}\delta(x'-x'_{a})\delta(y'-y'_{a})+m_{b}\delta(x'-x'_{b})\delta(y'-y'_{b})]\nonumber \\
\label{qi}
\end{eqnarray}
Here $\phi$ is the angle subtended by the object at the centre of mass (c. m.) with respect to the semi-major axis, $\tilde{r}$ is the distance between the objects, $m$ is the reduced mass of the system i.e. $m=m_{a}m_{b}/(m_{a}+m_{b})$ and c. m. denotes centre of mass of the system.

Using the above definitions, we get the simplified expression for the energy loss from a binary pulsar system as
\begin{eqnarray}
\dot{E}&=& -\frac{G m^{2}}{5c^{5}2\pi P}\left(2\pi l^{5}\frac{32}{a^{6}}\left(1+\frac{73}{24}e^{2}+\frac{37}{96}e^{4}\right)(1-e^{2})^{-6}\right) \nonumber \\
&-&\frac{G m^{2}}{5c^{5}2\pi P}\left(2L^{2}\frac{l^{7}1024\pi}{4c^{2}a^{10}}\left(1+\frac{23681}{1536}e^{2}+\frac{6071}{192}e^{4}+\frac{2201}{192}e^{6}+\frac{397}{1024}e^{8}\right)(1-e^{2})^{-10}\right)
\label{q1} 
\end{eqnarray}
where $l^{2}=aGM(1-e^2)$ ($M=m_{a}+m_{b}$), for elliptic systems.
 The steps leading to Eq. $\eqref{q1}$ have been shown in detail in  Appendix \ref{C}.

The energy loss given by Eq. $\eqref{q1}$ leads to the shrinkage of the orbit, resulting in a decay of the orbital period. In order to relate the energy loss with the orbital period decay, we consider the total energy of the system
\begin{equation}
E=\frac{m \dot{\tilde{r}}^2}{2} + \frac{l^2}{2 m \tilde{r}^2}-\frac{G m_{a}m_{b}}{\tilde{r}}
\label{r1}
\end{equation}
For an orbit with semi-major axis $a$, the total energy is given by (see, for example, Eq. $10$a 
in   \cite{blanchet} and Eq. $5.2$ in \cite{peters}),
\begin{equation}
E=-\frac{G m_{a}m_{b}}{2 a}
\label{r2}
\end{equation}
In order to express $a$ in terms of $P$, we use Kepler's third law: 
\begin{equation}
a^3 =\frac{GM}{4\pi^2}P^2
\label{r4}
\end{equation}
where $M=m_a+m_b$ is the total mass of the system. Using this relation and differentiating with respect to time we get a general relation between orbital period decay and energy loss:
%\begin{equation}
%\dot{E}=\frac{M_{1}M_{2}G}{3aP}\dot{P}\left(1-\frac{2\epsilon}{1-\epsilon}\frac{a^2}{L^2}\right)
%\end{equation}
\begin{equation}
\frac{\dot{P}}{P}=-\dot{E}\frac{3}{Gm_{a}m_{b}}a
\label{r5}
\end{equation}
For a given binary system, one can calculate the energy radiated using Eq. $\eqref{q1}$ with the known orbital parameters and equate  that energy loss to the change in potential energy given by Eq. $\eqref{r5}$ and hence calculate the change in orbital period.

\section{Rate of periastron advance}
\label{periastron}
We once again consider a two body system. The relative acceleration of the binary system is then expanded upto various Post-Newtonian (PN) orders. The 1PN correction terms give the periastron advance for an eccentric orbit \cite{Weinberg,will,will1}. It has already been shown in one of the earlier works on this model \cite{priti}, that the solution to the modified field equations in the weak field limit is given by the Yukawa potential. Hence if we expand the acceleration upto various PN orders, we expect that the lowest order contribution from Fourth Order Gravity will come due to the Yukawa kind of modification of the Newtonian potential. We expect that any other contributions/modifications coming from this model will be of higher order and much smaller in magnitude, and hence can be neglected (verification of this assumption for our model has been shown in the following section). Keeping the above facts in mind, we get the expression for the rate of change of periastron advance \cite{deng,will,will1} as
\begin{equation}
\langle\dot{\omega}\rangle=\frac{(2+2\gamma-\beta)GMn}{c^2 p}-\frac12\frac{n p^{2}}{L^{2}}\exp(-p/L)
\end{equation}
where $p=a(1-e^{2}), n=\sqrt{GM/a^{3}}$, $\gamma$ and $\beta$ are the PN parameters. The first term comes from PN approximation to GR and the second term corresponds to the contribution from Yukawa kind of a modification to Newtonian potential. Taking the values of the PN parameters i.e. $\gamma\approx\beta\approx 1$, which are the predicted values for GR and other modified gravity theories like STVG (Scalar Tensor Vector Gravity) \cite{will,will1}, we get
\begin{equation}
\langle\dot{\omega}\rangle=\frac{3GMn}{c^2 p}-\frac12\frac{n p^{2}}{L^{2}}\exp(-p/L)
\label{w1}
\end{equation}

\section{Results and Conclusions}
\label{constraints}
In this section, using the observations and predictions of GR for Hulse-Taylor binary pulsar, we obtain bounds on $L$. In general, in modified gravity theories, one gets a dipole radiation apart from the usual quadrupole radiation for binary systems with widely different primary and companion masses (e.g. white dwarf-neutron star system) \cite{will,will1}. In this work, we have considered neutron star-neutron star system for which the dipole radiation is negligible. For a detailed calculation of this claim, one may refer to \cite{Zhang:2017srh} where the authors have studied the dipole radiation for a model very similar to ours (See the Green's function in Eq. (56) of \cite{Zhang:2017srh}). The dipole contribution (relative to GR) has been given in Eq. (83) of the above reference, and is given by
\begin{equation}
\mathcal{A}_{dipole} = \frac{5}{192}\left(\frac{P}{2 \pi G M}\right)^{2/3} \mathcal{E}_d^2
\end{equation} 
where $\mathcal{E}_d = \epsilon_a -\epsilon_b$, $\epsilon_a,\ \epsilon_b$ being the inverse of the surface gravity of the neutron stars.

For the Hulse-Taylor pulsar, we get, $\mathcal{E}_d \approx 4.5 \times 10^{-18} s^2m^{-2}$. This gives the dipole contribution as $\mathcal{A}_{dipole} \sim 10^{-48}$ which is very weak in comparison to the quadrupole contribution relative to GR, which, in our case, is $\sim 0.03$. Hence we can neglect the dipole radiation for the Hulse-Taylor pulsar.

Below in Table \ref{tablea} we give the orbital parameters for the Hulse-Taylor binary pulsar system \cite{hulse,weisberg}.
\begin{table}[H]
\begin{center}
{\bf {\large Orbital Parameters for the Binary Pulsar B1913+16}}
\end{center}
\centering
\begin{tabular}{l l}
\hline\hline
$ {\rm {\bf Orbital \ parameters}} $ \quad \quad \quad \quad \quad \quad \quad \quad \quad \quad & $ {\rm {\bf Measured \ values}} $ \quad \quad \quad  \\ [1ex]
\hline\hline
${\rm eccentricity} \ e$ & $0.6171338$ \\ 
${\rm orbital\ period}\ P({\rm days})$ &  $0.322997448911$\\ 
${\rm mass\ function}\ f_m$ &  $0.13 M_{\odot}$ \\
${\rm orbital\ period\ decay}\ \dot{P} $ &  $ (-2.423 \pm 0.07) \times 10^{-12}$ \\ 
${\rm periastron\ advance}\ \dot{\omega}({\rm deg/yr}) $ &  $4.226598 \pm 0.0002$ \\[1ex]
\hline\hline 
\end{tabular}
\caption{Orbital parameters for PSR B1913+16 (Hulse-Taylor binary pulsar) \cite{weisberg}}
\label{tablea}
\end{table}

\bigskip

\noindent We can now obtain the constraints arising from period decay and periastron advance.
%\newpage
\bigskip

\noindent{\bf Bound from period decay}: 

\smallskip

\noindent Inserting $\eqref{q1}$ in $\eqref{r5}$, we get
\begin{equation}
\dot{P}= -\frac{3a P}{G m_a m_b}(\dot{E}_{GR}+ \delta\dot{E})
\end{equation}
where $\delta\dot{E}$ is the correction to the energy loss term due to fourth order modifications and is given by,
\begin{equation}
\delta\dot{E}= \frac{G m^{2}}{5c^{5}2\pi P}\left(2L^{2}\frac{l^{7}1024\pi}{4c^{2}a^{10}}\left(1+\frac{23681}{1536}e^{2}+\frac{6071}{192}e^{4}+\frac{2201}{192}e^{6}+\frac{397}{1024}e^{8}\right)(1-e^{2})^{-10}\right)
\end{equation}
To match with observations, this extra contribution should lie well within the error bars of $\dot{P}$ obtained from observations. 
To get a bound on $L$, we equate the error $\delta \dot{P}$ with the fourth order contribution,
\begin{equation}
\delta \dot{P} = \frac{3}{m_a m_b}\left[\frac{512 m^2 L^2 l^7}{10 c^7 a^{9}}\left(1+\frac{23681}{1536}e^2+\frac{6071}{192}e^4+ \frac{2201}{192}e^6 + \frac{397}{1024}e^8\right)(1-e^2)^{-10}\right]
\end{equation}
Using observational data ($\delta \dot{P} = 0.07 \times 10^{-12}$) from Table \ref{tablea}, with $m_a=1.41 M_{\odot}$ and $m_b=1.42 M_{\odot}$, we find that for the correction to be  within the error bar, $L \leq 4.187 \times 10^{10}$ m. 

\bigskip

%\paragraph{}

\noindent{\bf Bound from periastron advance}:

\smallskip

Similarly, equating the correction part due to Fourth Order Gravity with the error in the periastron advance for Hulse-Taylor pulsar, we get,
\begin{equation}
\delta \dot{\omega} =\frac{n p^2}{2 L^2}\exp(-p/L)
\end{equation}
Using observational data ($\delta \dot{\omega} = 0.0002$ deg/yr) from Table \ref{tablea}, we find that for the correction to be  within the error bar, either $L \geq 3.856 \times 10^{13}$ m or $L \leq 4.42 \times 10^{7}$ m . 

The lower bound ($L \geq 3.856 \times 10^{13}$) contradicts the bound from period dacay obtained earlier, hence we discard this. The upper bound i.e. $L \leq 4.42 \times 10^{7}$ m satisfies the period decay data also. We now need to verify that such low $L$ ($L \ll a; \; a\sim 10^{9}$ m) does not modify the Keplerian orbit. We start with the modified Poisson's equation in the weak field limit:
\begin{equation}
L^2\nabla^4 \phi - \nabla^2 \phi = -4 \pi G \rho
\end{equation}
The corresponding Green's function obtained from the above equation is
\begin{equation}
G(r)= -\frac{1}{r}+ \frac{e^{-r/L}}{r}
\end{equation}
When the source is a point mass with mass $M$, the solution of the potential can be found as,
\begin{equation}
\phi(r)= -\frac{G M}{r}+ \frac{G M e^{-r/L}}{r}
\end{equation}
Taking the limit $L \ll r$, we get,
\begin{equation}
\phi(r)= -\frac{G M}{r}
\end{equation}
which is the Newtonian potential as expected. Thus the limit $L \leq 4.42 \times 10^{7}$ does not modify the Keplerian orbit.

This bound also justifies our assumption of neglecting higher order PN terms coming from the modifications. For example, taking the upper bound i.e. $L\sim 10^7$ m, the lowest order contribution to the periastron advance due to the modification comes out to be $\sim 10^{40}$ times smaller than the contribution from GR. Hence, any higher order corrections can be safely discarded.

Thus, comparing the two observations (period decay and periastron advance), we obtain a common region of overlap i.e. $L \leq 4.42 \times 10^7$ m which satisfies both period decay as well as periastron advance data of the Hulse-Taylor pulsar.

\acknowledgments
We would like to thank A. Gopakumar for helpful discussions. 
\newpage

\centerline{\bf APPENDIX}

\appendix
\section{Perturbing the metric upto second order}
\label{A}
Expanding the metric upto second order, we can write,
\begin{equation}
g_{\mu\nu}=\eta_{\mu\nu}+h_{\mu\nu}+h^{(2)}_{\mu\nu}
\end{equation}

The Christoffel symbols can be written as
\begin{equation}
\Gamma = \Gamma^{(1)} + \Gamma^{(2)}  \quad \quad \quad {\rm since} \quad \Gamma^{(0)}=0
\end{equation}
The Ricci tensor will also be expanded upto second order,
\begin{equation}
R_{\mu\nu} = R_{\mu\nu}^{(1)} + R_{\mu\nu}^{(2)}
\label{e5}
\end{equation}
Putting this in Eq. $\eqref{e02}$, we get the energy-momentum carried by the gravitational wave,
\begin{equation}
t_{\mu\nu}= -\frac{1}{\kappa}\left(R^{(2)}_{\mu\nu}-\frac{1}{2}\eta_{\mu\nu} R^{(2)}-  L^2 (\nabla^{\alpha}\nabla^{\beta}R_{\mu\alpha\nu\beta})^{(2)}\right)
\label{e6}
\end{equation} 
  
The second order correction to the Ricci tensor is \cite{Carroll}
\begin{equation}
\begin{split}
R^{(2)}_{\mu\nu} &= \frac{1}{2}h^{\rho\sigma}\partial_{\mu}\partial_{\nu}h_{\rho\sigma}- h^{\rho\sigma}\partial_{\rho}\partial_{(\mu}h_{\nu)\sigma}+\frac14(\partial_{\mu}h_{\rho\sigma})\partial_{\nu}h^{\rho\sigma}\\
&+\partial^{\sigma}h^{\rho}_{\nu}\partial_{[\sigma}h_{\rho]\mu} + \frac12 \partial_{\sigma}(h^{\rho\sigma}\partial_{\rho}h_{\mu\nu})- \frac14(\partial_{\rho}h_{\mu\nu})\partial^{\rho}h\\
& - (\partial_{\sigma}h^{\rho\sigma}-\frac12\partial^{\rho}h)\partial_{(\mu}h_{\nu)\rho}\\
\end{split}
\label{e7}
\end{equation}
Using the harmonic gauge condition and the TT gauge, we can make the last two terms vanish. 

Now we calculate the last term of Eq. $\eqref{e6}$ i.e. the modifying term. Keeping upto second order, the double covariant derivative of the Riemann tensor is given by,
\begin{equation}
\begin{split}
(\nabla_{\alpha}\nabla_{\beta}R^{\mu\alpha\nu\beta})^{(2)}&= \partial_{\alpha}\partial_{\beta}R^{\mu\alpha\nu\beta (2)} + \Gamma^{\mu(1)}_{\sigma\alpha}\partial_{\beta}R^{\sigma\alpha\nu\beta (1)}\\
&+ \Gamma^{\alpha(1)}_{\sigma\alpha}\partial_{\beta}R^{\mu\sigma\nu\beta (1)}+ \Gamma^{\nu(1)}_{\sigma\alpha}\partial_{\beta}R^{\mu\alpha\sigma\beta (1)}+ \Gamma^{\beta(1)}_{\sigma\alpha}\partial_{\beta}R^{\mu\alpha\nu\sigma (1)}\\
&+ \partial_{\alpha}(\Gamma^{\mu(1)}_{\sigma\beta}R^{\sigma\alpha\nu\beta (1)}+ \cdots)\\
\end{split}
\label{e11}
\end{equation}
When we average over a large region (very large compared to the wavelength), then the fields vanish at the boundary and as a result, the total derivative terms go to zero. Finally performing the integration by parts and using the transverse-traceless condition, we get,
\begin{equation}
\left\langle (\nabla^{\alpha}\nabla^{\beta}R_{\mu\alpha\nu\beta})^{(2)} \right\rangle = \frac{1}{4}\left[\left\langle - h^{\sigma ,\alpha\beta}_{\nu}h_{\mu\sigma,\alpha\beta}- h^{\sigma\alpha ,\beta}_{,\nu}h_{\alpha\sigma,\mu\beta}\right\rangle\right]
\label{e12}
\end{equation}

Adding all these terms, we get $t_{\mu\nu}$ as
\begin{equation}
\begin{split}
\langle t_{\mu\nu}\rangle= -\frac{1}{\kappa}\Big\langle\Big(-\frac14(\partial_{\mu}h_{\rho\sigma})(\partial_{\nu}h^{\rho\sigma})-\frac{1}{2}\eta^{\rho\lambda}(\Box h_{\rho\nu})h_{\lambda\mu}&+\frac{L^2}{2}\eta_{\mu\nu}\eta^{\delta\omega}\left(-\frac14(\Box^2h_{\delta\sigma})h^{\sigma}_{\omega}-\frac14 h_{\sigma\beta,\alpha\delta}h^{\beta\sigma,\alpha}_{,\omega}\right)\\
&+\frac{L^2}{4}\left((\Box^2 h_{\nu\sigma})h^\sigma_\mu+h_{\sigma\beta,\nu\alpha}h^{\beta\sigma,\alpha}_{,\mu}\right)\Big)\Big\rangle
\end{split}
\end{equation}

\section{Evaluating the quadrupole integrals}
\label{B}
From Eq. $\eqref{s12}$, we can write 
\begin{equation}
h_{rr}=\frac{2G}{L}\int_{0}^{\infty}\frac{\mathcal{J}_{1}(s)}{\sqrt{s^2+\chi^2}}\ddot{\bar{Q}}_{rr}(t'_r)ds
\label{el3}
\end{equation}
 We calculate the derivatives of $h_{rr}$ as follows :

\begin{equation}
\partial_0 h_{rr}=\frac{2G}{L}\int_{0}^{\infty}\frac{\mathcal{J}_{1}(s)}{\sqrt{s^2+\chi^2}}\dddot{\bar{Q}}_{rr}(t'_r)ds
\end{equation}
\begin{equation}\begin{split}
\partial_r h_{rr}=-2G\left(\frac{1}{L^2}\int_0^\infty\frac{\chi\mathcal{J}_{1}(s)}{(s^2+\chi^2)^{3/2}}\ddot{\bar{Q}}_{rr}(t'_r)ds +\frac{1}{L}\int_0^\infty\frac{\chi\mathcal{J}_{1}(s)}{(s^2+\chi^2)}\dddot{\bar{Q}}_{rr}(t'_r)ds\right)
\end{split}
\end{equation}
\begin{equation}\begin{split}
\partial_r\partial_r h_{rr}&=-2G\Big(\frac{1}{L^3}\int_0^\infty \frac{\mathcal{J}_1(s)}{(s^2+\chi^2)^{3/2}}\ddot{\bar{Q}}_{rr}(t'_{r}) ds - \frac{3}{L^3}\int_0^\infty \frac{\chi^2\mathcal{J}_1(s)}{(s^2+\chi^2)^{5/2}}\ddot{\bar{Q}}_{rr}(t'_{r}) ds\\
 &+ \frac{1}{L^2}\int_0^\infty \frac{\mathcal{J}_1(s)}{s^2+\chi^2}\dddot{\bar{Q}}_{rr}(t'_{r}) ds - \frac{3}{L^2}\int_0^\infty \frac{\chi^2\mathcal{J}_1(s)}{(s^2+\chi^2)^{2}}\dddot{\bar{Q}}_{rr}(t'_{r}) ds - \frac{1}{L}\int_0^\infty \frac{\chi^2\mathcal{J}_1(s)}{(s^2+x^2)^{3/2}}\ddddot{\bar{Q}}_{rr}(t'_{r}) ds\Big)
\end{split}
\end{equation}
\begin{equation}
\partial_0\partial_r h_{rr}=-2G\left(\frac{1}{L^2}\int_0^\infty \frac{\chi\mathcal{J}_1(s)}{(s^2+\chi^2)^{3/2}}\dddot{\bar{Q}}_{rr}(t'_{r}) ds + \frac{1}{L}\int_0^\infty \frac{\chi\mathcal{J}_1(s)}{s^2+\chi^2}\ddddot{\bar{Q}}_{rr}(t'_{r}) ds\right)
\end{equation}
\begin{equation}
\partial_0\partial_0 h_{rr}=\frac{2G}{L}\int_{0}^{\infty}\frac{\mathcal{J}_{1}(s)}{\sqrt{s^2+\chi^2}}\ddddot{\bar{Q}}_{rr}(t'_r)ds
\end{equation}

Using the above relations and taking the far-field limit i.e. $\chi \to \infty$, we get the power radiated by the source in terms of its quadrupole moment as
\begin{eqnarray}
\dot{E} &&= \int d\Omega \frac{1}{4\kappa}\Big\langle -4G^2\int_0^\infty \frac{\chi \mathcal{J}_1(s)}{\sqrt{s^2+\chi^2}}ds \int_0^\infty \frac{\chi^2\mathcal{J}_1(s)}{s^2+\chi^2}ds \dddot{\bar{Q}}_{rr}(t'_{r})\dddot{\bar{Q}}^{rr}(t'_{r}) \nonumber \\ 
&&~~~~~~~~~~~~~~~~~ - L^2\Big(-4G^2\int_0^\infty \frac{\chi\mathcal{J}_1(s)}{s^2+\chi^2}ds \int_0^\infty \frac{\chi^4\mathcal{J}_1(s)}{(s^2+\chi^2)^{3/2}}ds \ddddot{\bar{Q}}_{rr}(t'_{r})\ddddot{\bar{Q}}^{rr}(t'_{r}) \nonumber \\&&~~~~~~~~~~~~~~~~~~~~~~~~~ +4G^2\int_0^\infty \frac{\chi^3\mathcal{J}_1(s)}{s^2+\chi^2}ds \int_0^\infty \frac{\mathcal{J}_1(s)}{\sqrt{s^2+\chi^2}}ds \ddddot{\bar{Q}}_{rr}(t'_{r})\ddddot{\bar{Q}}^{rr}(t'_{r})\Big)\Big\rangle \\
&& = -4G^2\int d\Omega \frac{1}{4\kappa}\Big\langle (\dddot{\bar{Q}}_{rr}\dddot{\bar{Q}}^{rr}+2L^2\ddddot{\bar{Q}}_{rr}\ddddot{\bar{Q}}^{rr})\Big\rangle 
\label{e15}
\end{eqnarray}
where we have used the following integration results \cite{HTFvol2}:
\begin{eqnarray}
\int_0^\infty \frac{\chi^2\mathcal{J}_1(s)}{s^2+\chi^2}ds&& = \frac{\chi}{2}[\mathcal{S}_1(\chi)+\pi E_1(\chi)+\pi \it{Y}_1(\chi)]\nonumber \\
&& = \frac{1}{2}(2-\frac{1}{\chi}+\frac{1}{\chi^2}) \quad\quad\quad \rm{(neglecting\ higher\ order\ terms\ as\ we\ are\ in\ the\ large}\ \chi\ \rm{regime)}\nonumber \\
\int_0^\infty \frac{\mathcal{J}_1(s)}{\sqrt{s^2+\chi^2}}ds && = \mathcal{I}_{1/2}(\chi/2)\mathcal{K}_{1/2}(\chi/2)
\end{eqnarray}
Here $\mathcal{J}_n$ and $\it{Y}_n$ are the Bessel's function of the first kind and second kind respectively, $E_n$ is the Weber's function and $\mathcal{S}_n$ is Schl\"{a}fli's polynomial. $\mathcal{I}_n$ and $\mathcal{K}_n$ are the modified Bessel's function of the first and second kind respectively.
  
We want to transform the above expression to the original form $Q_{ij}$ by extracting the transverse part \cite{mtw}. For this purpose, we use the projection operator as already defined in section \ref{source}. Here, we show some steps for the calculation of the first term of the above equation. The steps for calculating the second term are similar. Using Eq. $\eqref{s11.1}$, we can write,
\begin{equation}
\begin{split}
\dddot{\bar{Q}}^{ij} \dddot{\bar{Q}}_{ij} &= (P^a_i P^b_j-\frac12 P^{ab}P_{ij})(P^i_c P^j_d-\frac12 P_{cd}P^{ij})\dddot{Q}^{ab} \dddot{Q}_{cd}\\
&= (P^a_c P^b_d-\frac12P_{cd}P^{ab})\dddot{Q}^{ij} \dddot{Q}_{ij}\\
\end{split}
\label{TT}
\end{equation} 
where we have used $P^a_i P^i_b = P^a_b$. Using the projection operator defined earlier, we get,
\begin{equation}
\begin{split}
&P^a_c P^b_d = \delta^a_c \delta^b_d - \delta^a_c \frac{x^b x_d}{r^2} - \delta^b_d \frac{x^a x_c}{r^2} + \frac{x^a x_c x^b x_d}{r^4}\\
& P_{cd}P^{ab} = \delta_{cd}\delta^{ab} - \delta^{ab} \frac{x_c x_d}{r^2} - \delta_{cd} \frac{x^a x^b}{r^2} + \frac{x^a x^b x_c x_d}{r^4}\\
\end{split}
\label{el6}
\end{equation}
Putting all these together in Eq. $\eqref{TT}$ , we can write
\begin{equation}
\dddot{\bar{Q}}^{ij} \dddot{\bar{Q}}_{ij} = \dddot{Q}^{ab} \dddot{Q}_{ab} -\frac{x^b x^d}{r^2} \dddot{Q}_{ab} \dddot{Q}^a_d - \frac{x^a x^c}{r^2} \dddot{Q}_{ab} \dddot{Q}^b_c + \frac12 \frac{x^a x^b x^c x^d}{r^4} \dddot{Q}_{ab} \dddot{Q}_{cd}
\label{el7}
\end{equation}

In the next step, we will have to evaluate the integrals with respect to the solid angle $d\Omega$ over the surface of the sphere. To do this, we use some standard integrals as given below :
\begin{equation}
\begin{split}
& \int d\Omega = 4 \pi \\
& \int d\Omega x^a x^b = \frac43 \pi r^2 \eta^{ab}\\
& \int d\Omega x^a x^b x^c x^d = \frac{4}{15} \pi r^4 (\eta^{ab}\eta^{cd}+\eta^{ac}\eta^{bd}+\eta^{ad}\eta^{bc})\\
\end{split}
\label{el8}
\end{equation}

With these integrals and using the traceless property of $Q_{ij}$, we finally have
\begin{equation}
\int \dddot{\bar{Q}}^{ij} \dddot{\bar{Q}}_{ij} d\Omega = \frac{8\pi}{5} \dddot{Q}^{ij} \dddot{Q}_{ij}
\label{el9}
\end{equation}
We can repeat the above steps for the second term also and get
\begin{equation}
\int \ddddot{\bar{Q}}^{ij} \ddddot{\bar{Q}}_{ij} d\Omega = \frac{8\pi}{5} \ddddot{Q}^{ij} \ddddot{Q}_{ij}
\label{el9.1}
\end{equation}
Using these results in Eq. $\eqref{e15}$, we finally get (including $c$),
\begin{equation}
\dot{E}=-\frac{G}{5c^5}\Big\langle (\dddot{Q}_{rr}\dddot{Q}^{rr}+\frac{2L^2}{c^2}\ddddot{Q}_{rr}\ddddot{Q}^{rr})\Big\rangle
\end{equation}
\section{Expressing energy loss in terms of orbital parameters}
\label{C}
Using Eqs. $\eqref{qi}$, we calculate the various components of the quadrupole moment as,

\begin{eqnarray}
I_{x'x'}&=&\frac12(\cos(2\phi)+1)\tilde{r}^{2}m \quad ;\quad I_{y'y'}=\frac12(1-\cos(2\phi))\tilde{r}^{2}m;\nonumber \\
I_{x'y'}&=&I_{y'x'}=\frac12\sin(2\phi)\tilde{r}^{2}m\quad ;\quad \frac13 \delta_{\mu\nu}\delta^{ij}I_{ij}=\frac13 \delta_{\mu\nu}\tilde{r}^{2}m
\end{eqnarray}
Using the above results we get the different non-zero components of the reduced quadrupole momentum as
\begin{eqnarray}
Q_{x'x'}&=&\frac12 \tilde{r}^{2}m(\cos(2\phi)+\frac13)\quad ;\quad Q_{x'y'}=\frac12 \tilde{r}^{2}m\sin(2\phi)\nonumber \\
Q_{y'x'}&=&\frac12 \tilde{r}^{2}m\sin(2\phi)\quad \quad \quad \quad ;\quad  Q_{y'y'}=-\frac12 \tilde{r}^{2}m\cos(2\phi)\nonumber \\
Q_{z'z'}&=&-\frac13 \tilde{r}^{2}m
\end{eqnarray}
In order to take the third time derivative of the reduced quadrupole moment, we define a new quantity known as the angular momentum per unit mass given by $l=\tilde{r}^{2}\dot{\phi}$, which is a conserved quantity, and hence can be taken out of the integral. Also, in order to make the calculations easier, we define $u=\tilde{r}^{-1}$, $d/dt=lu^{2}d/d\phi$.

Using the above results, we get the third time derivative of the reduced quadrupole moment as
\begin{equation}
\dddot{Q}_{ij}=2u^{2}l^{3}(u^{2}(u'u''-uu''')Q_{ij}-2u^{3}(u''+u)Q'_{ij})
\end{equation} 
where $'$ denotes derivative with respect to $\phi$.

In order to take the average over time, we use the following results
\begin{eqnarray}
Tr(\dddot{Q}^{2})&=& \frac{m^{2}l^{5}}{2\pi P}\int_{0}^{2\pi} d\phi u^{-2}\left[\frac{8}{3}u^{4}(u'u''-uu''')^{2}+32 u^{6}(u''+u)^{2}\right]; \\
Tr(\ddddot{Q}^{2})&=& \frac{m^{2}l^{7}}{4\pi P}\int_{0}^{2\pi}d\phi u^{4}\left[\frac{32}{3}(-2u^{2}u'u'''-u^{3}u''''+2uu'^{2}u''+u^{2}u''^{2})^{2}+32(3u^{3}u'''+5u^{2}u'u''+8u^{3}u')^{2}\right] \nonumber \\ &+& \frac{m^{2}l^{7}}{4\pi P}\int_{0}^{2\pi}d\phi u^{4}\left[512(u^{3}u''+u^{4})^{2}-128(-2u^{2}u'u'''-u^{3}u''''+2uu'^{2}u''+u^{2}u''^{2})(u^{3}u''+u^{4})\right]\nonumber \\
\end{eqnarray}
Using the solution for $u$ given by $u=a^{-1}(1-e^{2})^{-1}(1-e\cos(\phi))$, we get the simplified expression for energy loss,
\begin{eqnarray}
\dot{E}&=& -\frac{G m^{2}}{5c^{5}2\pi P}\left(2\pi l^{5}\frac{32}{a^{6}}\left(1+\frac{73}{24}e^{2}+\frac{37}{96}e^{4}\right)(1-e^{2})^{-6}\right) \nonumber \\
&-&\frac{G m^{2}}{5c^{5}2\pi P}\left(2L^{2}\frac{l^{7}1024\pi}{4c^{2}a^{10}}\left(1+\frac{23681}{1536}e^{2}+\frac{6071}{192}e^{4}+\frac{2201}{192}e^{6}+\frac{397}{1024}e^{8}\right)(1-e^{2})^{-10}\right)
\end{eqnarray}

\end{document}